\documentstyle[prd,aps,floats,epsf]{revtex}

\begin{document}
\title{\Large {\bf The Standard Model of Particle Physics}}
\author{\large Mary K. Gaillard$^1$, 
Paul D. Grannis$^2$, and Frank J. Sciulli$^3$}
\vskip 3mm
\address{$^1$University of California, Berkeley,
$^2$State University of New York, Stony Brook,
$^3$Columbia University}
\maketitle

\begin{abstract}
Particle physics has evolved a coherent
model that characterizes forces and particles at the most 
elementary level.   This Standard Model, 
built from many theoretical and experimental studies, is in excellent
accord with almost all current data.   However, there are many hints that
it is but an approximation to a yet more fundamental theory.   We trace 
the development of the Standard Model and indicate the reasons for believing
that it is incomplete.
\end{abstract}

\vskip 5mm

\centerline{Nov. 20, 1998}
\centerline{\sl (To be published in Reviews of Modern Physics)}

\large
\baselineskip=16pt
\section{Introduction: A Bird's Eye View of the Standard Model}

Over the past three decades a compelling case has emerged for the
now widely accepted Standard Model of elementary particles and forces.
A `Standard Model' is a theoretical framework built from observation that
predicts and correlates new data.    The Mendeleev table of elements was 
an early example in chemistry; from
    the periodic table one could predict the properties of many hitherto
    unstudied elements and compounds.    Nonrelativistic quantum theory is
    another Standard Model that has correlated the results of countless
experiments.   Like its precursors in other fields, the Standard Model (SM) of
    particle physics has been enormously successful in predicting a 
wide range of phenomena. And, just as ordinary quantum mechanics fails in the 
relativistic limit, we do not expect the SM to be 
valid at arbitrarily short distances. However its remarkable success strongly 
suggests that the SM will remain an excellent approximation to
nature at distance scales as small as 10$^{-18}$ m.

In the early 1960's particle physicists described nature in terms of four
distinct forces, characterized by widely different ranges and strengths as 
measured at a typical energy scale of 1 GeV.  The strong nuclear force has a 
range of about a fermi or $10^{-15}$ m. The weak force responsible for 
radioactive decay, with a range of $10^{-17}$ m, is about $10^{-5}$ times 
weaker at low energy. The electromagnetic force that governs much
of macroscopic physics has infinite range and strength 
determined by the fine structure constant, $\alpha\approx 10^{-2}$. 
The fourth force, gravity, also has infinite range and a low energy coupling
(about $10^{-38}$)  too weak to be observable in laboratory experiments. 
The achievement of the SM was the elaboration
of a unified description of the strong, weak and electromagnetic forces in
the language of quantum gauge field theories.  Moreover, the SM 
combines the weak and electromagnetic forces in a single electroweak gauge 
theory, reminiscent of Maxwell's unification of the seemingly distinct forces 
of electricity and magnetism.   

By mid-century, the electromagnetic force was well understood as a
renormalizable quantum field theory (QFT) known as quantum electrodynamics or 
QED, described in the preceeding article.
`Renormalizable' means that once a few parameters are determined by a limited 
set of measurements, the quantitative features  of interactions among charged
particles and photons can be calculated to arbitrary accuracy as a 
perturbative expansion in the fine structure constant.  QED has been
tested over an energy range from $10^{-16}$ eV 
to tens of  GeV, {\it i.e.} distances ranging from $10^{8}$ km to
$10^{-2}$ fm. In contrast, the nuclear force was characterized by
a coupling strength that precluded a perturbative expansion. 
Moreover, couplings involving higher spin states (resonances), that 
appeared to be on the same footing as nucleons and pions, could not be 
described by a renormalizable theory, nor could the weak interactions that were
attributed to the direct coupling of four fermions to one another.
In the ensuing years the search for renormalizable theories of strong and
weak interactions, coupled with experimental discoveries and attempts to 
interpret available data, led to the formulation of the SM, which 
has been experimentally verified to a high degree of accuracy over 
a broad range of energy and processes.

The SM is characterized in part by the spectrum of elementary 
fields shown in Table I. 
The matter fields are fermions and their anti-particles,
with half a unit of intrinsic angular 
momentum, or spin. There are three families of fermion fields that are 
identical in every attribute except their masses.  The first family includes 
the up ($u$) and down ($d$) quarks that are the constituents of nucleons as 
well as pions and other mesons responsible for nuclear binding. It also 
contains the electron and the neutrino emitted with a positron in nuclear $
\beta$-decay. The quarks of the other families are constituents of 
heavier short-lived particles; they and their companion charged leptons 
rapidly decay via the weak force to the quarks and leptons of the first family.

The spin-1 gauge bosons mediate interactions among fermions.  
In QED, interactions among electrically charged particles are due to the 
exchange of quanta of the electromagnetic field called photons ($\gamma$). The 
fact that the $\gamma$ is massless accounts for the long range of the 
electromagnetic force.  The strong force, quantum chromodynamics or QCD, is 
mediated by the exchange of massless gluons ($g$) between quarks that carry a 
quantum number called color.  In contrast to the electrically neutral photon, 
gluons (the quanta of the `chromo-magnetic' field) possess color 
charge and hence couple to one another.  As a consequence,
the color force between two colored particles increases 
in strength with increasing distance. Thus quarks and gluons cannot appear as 
free particles, but exist only inside composite particles, called hadrons, 
with no net color charge. Nucleons are composed of three 
quarks of different colors, resulting in `white' color-neutral 
states.  Mesons contain quark and anti-quark
pairs whose color charges cancel.  Since a gluon inside a nucleon cannot escape
its boundaries, the nuclear force is mediated by color-neutral bound states, 
accounting for its short range, characterized by the Compton wavelength of the 
lightest of these:  the $\pi$-meson.  

The even shorter range of the weak force is associated with the Compton 
wave-lengths of the charged $W$ and neutral $Z$ bosons that mediate it. Their
couplings to the `weak charges' of quarks and leptons are comparable in 
strength to the electromagnetic coupling.  When the weak interaction is 
measured 
over distances much larger than its range, its effects are averaged over the 
measurement area and hence suppressed in amplitude by a factor $(E/M_{W,Z})^2 
\approx (E/100$ GeV)$^2$, 
where $E$ is the characteristic energy transfer in the 
measurement.  Because the $W$ particles carry electric charge they must couple
to the $\gamma$, implying a gauge theory  that unites the weak and 
electromagnetic interactions, similar to QCD in that the gauge
particles are self-coupled.  In distinction to $\gamma$'s 
and gluons, $W$'s couple only to left-handed fermions 
(with spin oriented opposite to the direction of motion).

The SM is further characterized by a high degree of symmetry.  For 
example, one cannot perform an
experiment that would distinguish the color of the quarks involved.  If the
symmetries of the SM couplings were fully
respected in nature, we would not distinguish an
electron from a neutrino or a proton from a neutron;  their detectable 
differences are attributed to 
`spontaneous' breaking of the symmetry.  Just as the spherical symmetry
of the earth is broken to a cylindrical symmetry by the earth's magnetic field,
a field permeating all space, called the Higgs field, is invoked to explain the
observation that the symmetries of the electroweak theory are broken 
to the residual gauge symmetry of QED.  Particles that interact with the 
Higgs field cannot propagate at the speed of light, and acquire masses, in
analogy to the index of refraction that slows a photon traversing matter.  
Particles that do not interact with the Higgs field --- 
the photon, gluons and possibly neutrinos -- remain massless.  
Fermion couplings to the Higgs field not only determine their masses; they 
induce a misalignment of quark mass eigenstates with respect to the 
eigenstates of the weak charges, thereby allowing all fermions of heavy 
families to decay to lighter ones. 
These couplings provide the only mechanism within the SM that 
can account for the observed violation of CP, that is,
invariance of the laws of nature under mirror reflection (parity P) and  
the interchange of particles with their anti-particles (charge
conjugation C).

The origin of the Higgs field has not yet been determined. However our very 
understanding of the SM implies that physics associated with 
electroweak symmetry breaking (ESB) must become manifest at
energies of present colliders or at the LHC under construction.  There is
strong reason, stemming from the quantum instability of scalar masses,
to believe that this physics will point to modifications of the
theory.  One shortcoming of the SM is its failure to 
accommodate gravity, for which there is no renormalizable QFT
because the quantum of the gravitational field has two units of 
spin.  Recent theoretical progress suggests that quantum gravity can be
formulated only in terms of extended objects like strings and membranes, with
dimensions of order of the Planck length $~ 10^{-35}$ m. 
Experiments probing higher energies and shorter distances may reveal clues
connecting SM physics to gravity, and may shed light on other
questions that it leaves unanswered.  In the following we
trace the steps that led to the formulation of the SM, describe the
experiments that have confirmed it, and discuss some outstanding unresolved
issues that suggest a more fundamental theory underlies the SM.

\section{The path to QCD}

The invention of the bubble chamber permitted the observation of a 
rich spectroscopy of hadron states. Attempts at their classification using group
theory, analogous to the introduction of isotopic spin as a classification 
scheme for nuclear states, culminated in the 
`Eightfold Way' based on the group SU(3), in which particles are ordered by 
their `flavor' quantum numbers: isotopic spin and strangeness.  This scheme 
was spectacularly confirmed by the discovery at Brookhaven Laboratory (BNL)
of the $\Omega^-$ particle, with
three units of strangeness, at the predicted mass.  It was subsequently 
realized that the spectrum of the Eightfold Way could be
understood if hadrons were composed of three types of quarks: $u$, $d$, and 
the strange quark $s$.  However the quark model presented a dilemma: 
each quark was attributed one half unit of spin, but Fermi statistics precluded
the existence of a state like the $\Omega^-$ composed of three strange quarks
with total spin ${3\over2}$. Three identical fermions with their spins aligned
cannot exist in an an $s$-wave ground state.  This paradox led to the
hypothesis that quarks possess an additional quantum number called color, a
conjecture supported by the observed rates for $\pi^0$ decay into 
$\gamma \gamma$ and $e^+e^-$ annihilation into hadrons, both of which require
three different quark types for each quark flavor.

A combination of experimental observations and theoretical analyses in the
1960's led to another important conclusion: pions behave like the Goldstone 
bosons of a spontaneously broken symmetry, called chiral symmetry. Massless
fermions have a conserved quantum number called chirality, equal to 
their helicity:  $+1(-1)$ for right(left)-handed fermions. The 
analysis  of pion scattering lengths and weak decays into pions 
strongly suggested that chiral symmetry is explicitly broken only by quark 
masses, which in turn implied that the underlying theory describing strong 
interactions among quarks must conserve quark
helicity -- just as QED conserves electron helicity.  This further implied 
that interactions among quarks must be mediated by the exchange of spin-1
particles. 

In the early 1970's, experimenters at the Stanford Linear Accelerator Center
(SLAC) analyzed the distributions in energy and angle of electrons scattered 
from nuclear targets in inelastic collisions with 
momentum transfer $Q^2$ $\approx 1$ GeV/c 
from the electron to the struck nucleon. The distributions 
they observed suggested that electrons interact via photon exchange with 
point-like objects called partons -- 
electrically charged particles much smaller than nucleons.  If the electrons 
were scattered by an extended object, {\it e.g.} a strongly interacting
nucleon with its electric charge spread out by a cloud of pions, the 
cross section would drop rapidly for values of momentum transfer 
greater than the inverse radius of the charge distribution.  Instead, the data
showed a `scale invariant' distribution:  a cross section equal to
the QED cross section up to a dimensionless function of kinematic variables, 
independent of the energy of the incident electron.  Neutrino scattering 
experiments at CERN and Fermilab (FNAL) yielded similar results.
Comparison of electron and neutrino data allowed a determination of
the average squared electric charge of the partons in the nucleon, and the
result was consistent with the interpretation that they are fractionally
charged quarks.
Subsequent experiments at SLAC showed that, at center-of-mass energies above 
about two  GeV, the final states in $e^+e^-$ annihilation into hadrons have a
two-jet configuration. The angular distribution of the jets with respect to the
beam, which depends on the spin of the final state particles, is
similar to that of the muons in an $\mu^+\mu^-$ final state,  providing 
direct evidence for spin-${1\over2}$ parton-like objects.  

\section{The path to the electroweak theory}

A major breakthrough in deciphering the structure of weak interactions was
the suggestion that they may not conserve parity, 
prompted by the observation of $K$-decay into both $2\pi$ 
and $3\pi$ final states with opposite parity. An intensive search for 
parity violation in other decays 
culminated in the establishment of the `universal $V-A$
interaction'.  Weak processes such as nuclear $\beta$-decay and muon decay 
arise from quartic couplings of fermions with negative chirality; thus only 
left-handed electrons and right-handed positrons 
are weakly coupled.  Inverse $\beta$-decay was observed in interactions 
induced by electron anti-neutrinos from reactor fluxes, and several years 
later the muon neutrino was demonstrated to be distinct from the electron
neutrino at the BNL AGS.

With the advent of the quark model, the predictions
of the universal $V-A$ interaction could be summarized by introducing a weak
interaction Hamiltonian density of the form
\begin{eqnarray}
~~~~H_w = &~&{G_F\over\sqrt{2}}J^\mu J^{\dag}_\mu ~~, \nonumber\\ 
J_{\mu} = &~&\bar{d}\gamma_\mu(1 - \gamma_5)u + 
            \bar{e}\gamma_\mu(1 - \gamma_5)\nu_e  
          + \bar{\mu}\gamma_\mu(1 - \gamma_5)\nu_\mu ~~, 
\end{eqnarray}
where $G_F$ is the Fermi coupling constant, $\gamma_\mu$ is a Dirac matrix and 
${1\over2}(1-\gamma_5)$ is the negative chirality projection operator.  However
(1) does not take into account the observed $\beta$-decays of strange 
particles.  Moreover, increasingly precise measurements, together
with an improved understanding of quantum QED corrections, showed that
the Fermi constant governing neutron $\beta$-decay is a few percent less than
the $\mu$ decay constant.  Both problems were resolved by the introduction of
the Cabibbo angle $\theta_c$ and the replacement 
$d\to d_c = d\,\cos\theta_c + s\,\sin\theta_c$ in (1).  Precision
measurements made possible by high energy beams of
hyperons (the strange counterparts of nucleons) at CERN and FNAL 
have confirmed in detail the predictions of this theory with $\sin\theta_c 
\approx 0.2$.

While the weak interactions maximally violate P and C, CP 
is an exact symmetry of the Hamiltonian (1). The discovery at BNL in 
1964 that CP is violated in neutral kaon decay to two pions at a level of 0.1\%
in amplitude could not be incorporated into the theory in any obvious way.
Another difficulty arose from quantum effects induced by the Hamiltonian (1)
that allow the annihilation of the anti-strange quark and the down quark in a 
neutral kaon. This annihilation can produce a $\mu^+\mu^-$ pair, 
resulting in the decay $K^0\to \mu^+\mu^-$, or a $\bar{d} s$ pair, inducing 
$K^0$-$\bar{K}^0$ mixing. 
To suppress processes like these to a level consistent with 
experimental observation, a fourth quark flavor called charm ($c$) was 
proposed, with the current density in (1) modified to read
\begin{eqnarray}
 J_\mu = &~& \bar{d}_c\gamma_\mu(1 - \gamma_5)u + 
            \bar{s}_c\gamma_\mu(1 - \gamma_5)c 
          + \bar{e}\gamma_\mu(1 - \gamma_5)\nu_e + 
           \bar{\mu}\gamma_\mu(1 - \gamma_5)\nu_\mu ~~, \nonumber\\
   s_c = &~& s\,\cos\theta_c - d\,\sin\theta_c ~~. 
\end{eqnarray}
With this modification, contributions from virtual $c\bar{c}$ pairs cancel
those from virtual $u\bar{u}$ pairs, up to effects dependent on the difference
between the $u$ and $c$ masses.  Comparison with experiment suggested that the
charmed quark mass should be no larger than a few  GeV. The narrow resonance
$J/\psi$ with mass of about 3 GeV, found in 1974 at BNL and SLAC, was 
ultimately identified as a $c\bar{c}$ bound state.

\section{The search for renormalizable theories}

In the 1960's the only known renormalizable theories were QED and the Yukawa
theory -- the interaction of spin-${1\over2}$ fermions via the exchange of 
spinless particles.  Both the chiral symmetry of the strong
interactions and the $V-A$ nature of the weak interactions suggested that all
forces except gravity 
are mediated by spin-1 particles, like the photon.  QED is renormalizable
because gauge invariance, which gives conservation of electric charge, also
ensures the cancellation of quantum corrections that would otherwise result in
infinitely large amplitudes.  Gauge invariance implies a
massless gauge particle and hence a long-range force.  Moreover the
mediator of weak interactions must carry electric charge and thus couple to
the photon, requiring its description within a Yang-Mills theory that is
characterized by self-coupled gauge bosons.

The important theoretical breakthrough of the early 1970's was the proof 
that Yang-Mills theories are renormalizable, and that 
renormalizability remains intact if gauge symmetry is spontaneously broken,
that is, if the Lagrangian is gauge invariant, but
the vacuum state and spectrum of particles are not. An example
is a ferromagnet for which the lowest energy configuration has
electron spins aligned; the direction of alignment spontaneously breaks 
the rotational invariance
of the laws of physics.  In QFT, the simplest way to induce
spontaneous symmetry breaking is the Higgs mechanism. 
A set of elementary scalars $\phi$ is introduced with a potential energy
density function $V(\phi)$ that is minimized at a value $<$$\phi$$>\ne
0$ and the vacuum energy is degenerate.  For example, the 
gauge invariant potential for an electrically charged scalar field $\phi =
|\phi|e^{i\theta}  $,
\begin{equation}
 V(|\phi|^2) = - \mu^2 |\phi|^2 + \lambda |\phi|^4, \end{equation} 
has its minimum at $\sqrt{2}$$<$$|\phi|$$> = \mu/\sqrt{\lambda} = v$, 
but is independent of 
the phase $\theta$. Nature's choice for $\theta$ spontaneously breaks the 
gauge symmetry.  Quantum excitations of $|\phi|$ about its vacuum value are
massive Higgs scalars: $m^2_H = 2\mu^2 = 2\lambda v^2.$
Quantum excitations around the vacuum value of $\theta$ cost no energy and
are massless, spinless particles called Goldstone bosons. They appear in the
physical spectrum as the longitudinally polarized spin states of 
gauge bosons that acquire masses through their couplings to the Higgs field. A 
gauge boson mass $m$ is determined by its coupling $g$ to the Higgs field and 
the vacuum value $v$.  Since gauge couplings are universal this also 
determines the Fermi constant $G$ for this toy model: 
$m = gv/2,\;G/\sqrt{2} = g^2/8m^2 = v^2/2$.  

The gauge theory of electroweak interactions entails four gauge bosons:
$W^{\pm 0}$ of SU(2) or weak isospin $\vec I_w$, with coupling constant $g =
e\sin\theta_w$, and $B^0$ of U(1) or weak hypercharge $Y_w = Q - I^3_w$, 
with coupling $g' =e\cos\theta_w$.  Symmetry breaking can be achieved by the 
introduction of an isodoublet of complex scalar fields $\phi = (\phi^+\; 
\phi^0)$, with a potential identical to (3) where $|\phi|^2 = |\phi^+|^2 + 
|\phi^0|^2$.  Minimization of the vacuum energy fixes $v = \sqrt{2}|\phi| = 
2^{1\over4}G_F^{1\over2} =$ 246 GeV, leaving three Goldstone
bosons that are eaten by three massive vector bosons: $W^{\pm}$ and $Z =
\cos\theta_w W^0 - \sin\theta_w B^0$, while the photon $\gamma = 
\cos\theta_w B^0 + \sin\theta_w W^0$ remains massless.  This theory predicted
neutrino-induced neutral current (NC) interactions of the type $\nu +$ atom 
$\to\nu +$ anything, mediated by $Z$ exchange. The weak mixing angle 
$\theta_w$ governs the dependence of NC couplings on fermion 
helicity and electric charge, and their interaction rates are
determined by the Fermi constant $G_F^Z$. The ratio 
$\rho = G_F^Z/G_F =  m^2_W/m_Z^2\cos^2\theta_w$, predicted to be 1, is the 
only measured parameter of the SM 
that probes the symmetry breaking mechanism.  Once the value of $\theta_w$ was
determined in neutrino experiments, the $W$ and $Z$ masses could be predicted:
$ m^2_W = m^2_Z\cos^2\theta_w = \sin^2\theta_w\pi\alpha/\sqrt{2}G_F.$

This model is not renormalizable with three quark flavors and four lepton
flavors because gauge 
invariance is broken at the quantum level unless the sum of electric charges 
of all fermions vanishes.  This is true for each family of fermions in Table I, 
and could be achieved by invoking the existence of the charmed quark, 
introduced in (2).  However, the discovery of charmed mesons 
($c\bar{u}$ and $c\bar{d}$ bound states) in 1976 
was quickly followed by the discovery 
of the $\tau$ lepton,
requiring a third full fermion family.  A third family had in fact been
anticipated by efforts to accommodate CP violation, which can arise from 
the misalignment between fermion gauge couplings and Higgs couplings provided
there are more than two fermion families.

\begin{table*}
\begin{tabular}{||ll|ll|ll|ll||ll||}
\multicolumn{4}{||c}{QUARKS: $S={1\over2}$}&\multicolumn{4}{|c}{LEPTONS: 
$S={1\over2}$}&\multicolumn{2}{||c||}{GAUGE BOSONS: $S = 1$}\\ \hline\hline
$Q = {2\over3}$ & $\qquad m$ & $Q = -{1\over3}$ & $\qquad m$ & $Q$=-1 & 
$\qquad m$ & $Q$=0& $\qquad m$  & quanta & $\qquad m$ \\ \hline\hline
$u_1$ $u_2$ $u_3$ & (2--8)$10^{-3}$ & $d_1$ $d_2$ $d_3$ & (5--15)$10^{-3}$ 
& $e$ & 5.11$\times10^{-4}$ & $\nu_e$ & $<1.5\times10^{-8}$ & 
$g_1\cdots g_8$ & $<$ a few $\times10^{-3}$ \\ \hline 
$c_1$ $c_2$ $c_3$ & 1.0--1.6 & $s_1$ $s_2$ $s_3$ & 0.1--0.3 & $\mu$ & 0.10566 
& $\nu_\mu$ & $<1.7\times10^{-4}$ & $\gamma$ & $<6\times10^{-25}$ \\ \hline 
$t_1$ $t_2$ $t_3$ & 173.8$\pm$5.0 & $b_1$ $b_2$ $b_3$ & 
4.1--4.5 & $\tau$ & 1.7770 & $\nu_\tau$ & $<1.8\times10^{-2}$ & 
$W^{\pm},Z^0$ & 80.39$\pm$0.06,91.187$\pm$0.002 \\ 
\end{tabular}
\vskip 2mm
\caption{Elementary particles of the SM: $S(\hbar)$ is 
spin, $Q(e)$ is electric charge, and $m($GeV/$c^2)$ is mass. Numerical
subscripts indicate the distinct color states of quarks and gluons.}
\end{table*}

Meanwhile, to understand the observed scaling behavior in deep inelastic 
scattering (DIS) of leptons from nucleons, theorists were searching for an
asymptotically free theory -- a theory in which couplings become weak at short
distance.  The charge distribution of a strongly interacting 
particle is spread out by quantum effects, while scaling showed that at large
momentum-transfer quarks behaved like noninteracting particles.
This could be understood if the strong coupling becomes weak at short 
distances, in contrast to electric charge or Yukawa couplings that decrease
with distance due to the screening effect of vacuum polarization.  QCD, with 
gauged SU(3) color charge, became the prime candidate for the strong force when 
it was discovered that Yang-Mills theories are asymptotically free: the
vacuum polarization from charged gauge bosons has the opposite sign from the
fermion contribution and is dominant if there are sufficiently few fermion
flavors.  This qualitatively explains quark and gluon
confinement:  the force between color-charged particles grows with the distance
between them, so they cannot be separated by a distance 
much larger than the size of
a hadron.  QCD interactions at short distance are
characterized by weak coupling, and can be calculated using perturbation theory
as in QED; their effects contribute measurable deviations from scale 
invariance that depend logarithmically on the momentum transfer.

The SM gauge group, $SU(3)\times SU(2)\times U(1)$, is 
characterized by three coupling
constants $g_3 = g_S, \; g_2 = g, \; g_1 = \sqrt{5/3}g'$, where 
$g_1$ is fixed by requiring the same normalization for all fermion currents.
Their measured values at low energy
satisfy $g_3>g_2>g_1$.  Like $g_3$, the coupling $g_2$ decreases with
increasing energy, but more slowly because there are fewer gauge bosons
contributing.  As in QED, the $U(1)$ coupling increases with energy. Vacuum
polarization effects calculated using the particle content of the SM
show that the three coupling constants are very nearly equal at an 
energy scale around $10^{16}$ GeV, providing a tantalizing hint of a more 
highly symmetric theory, embedding the SM 
interactions into a single force.  Particle masses also
depend on energy; the $b$ and $\tau$ masses become equal at a similar scale, 
suggesting a possibility of quark and lepton unification as
different charge states of a single field.  

\section{Brief Summary of the Standard Model elements}

The SM contains the set of elementary particles 
shown in Table I.   The forces operative in the particle domain
are the strong (QCD) interaction responsive to particles carrying color,
and the two pieces of the electroweak interaction responsive to 
particles carrying weak isospin and hypercharge.   The quarks come in
three experimentally indistinguishable colors and there are eight
colored gluons.  All quarks and leptons, and the 
$\gamma$, $W$ and $Z$ bosons, carry weak isospin.  In the strict view
of the SM, there are no right-handed neutrinos or left-handed anti-neutrinos.
As a consequence the simple Higgs mechanism described in section IV cannot
generate neutrino masses, which are posited to be zero.

In addition, the SM provides the quark mixing matrix
which gives the transformation from the basis of the strong interaction
charge $-{1\over 3}$ left-handed quark flavors to the mixtures which couple to
the electroweak current.  The elements of this matrix are fundamental
parameters of the SM.   A similar mixing may occur for the neutrino
flavors, and if accompanied by nonzero neutrino mass, would induce 
weak interaction flavor-changing phenomena that are outside the 
SM framework.

Finding the constituents of the SM spanned the first century of
the APS, starting with the discovery by Thomson of the electron in 1897.
Pauli in 1930 postulated the existence of the neutrino as the agent
of missing energy and angular momentum in $\beta$-decay; only in 1953 was the
neutrino found in experiments at reactors.
The muon was unexpectedly added from cosmic ray searches for the Yukawa
particle in 1936;  in 1962 its companion neutrino was found in the decays
of the pion.

The Eightfold Way classification of the hadrons in 1961 suggested the
possible existence of the three lightest quarks ($u$, $d$ and $s$), though
their physical reality was then regarded as
doubtful.  The observation of substructure
of the proton, and the 1974 observation of the 
$J/\psi$ meson interpreted as a $c \overline c$ bound state and mesons
with a single charm quark in 1976, cemented the reality of the first two
generations of quarks.  This state of affairs, with two symmetric generations of
leptons and quarks, was theoretically tenable and the particle
story very briefly seemed finished.

In 1976, the $\tau$ lepton was found in a 
SLAC experiment, breaking new ground into
the third generation of fermions.   The discovery of the $\Upsilon$ at FNAL
in 1979 was interpreted as the bound state of a new 
bottom ($b$) quark.   The neutrino
associated with the $\tau$ has not been directly observed, but indirect
measurements certify its existence beyond reasonable doubt.  The final step
was the discovery of the top ($t$) 
quark at FNAL in 1995.    Despite the completed
particle roster, there are fundamental questions remaining; chief among these
is the tremendous disparity of the matter particle masses, ranging from
the nearly massless neutrinos, the 0.5 MeV electron 
and few MeV $u$ and $d$ quarks,
to the top quark whose mass is nearly 200 GeV.   Even the taxonomy of particles
hints at unresolved fundamental questions!

The gauge particle discoveries are also complete.   The photon was inferred from
the arguments of Planck, Einstein and Compton early in this century.
The carriers of the weak interaction, the $W$ and $Z$ bosons, were postulated
to correct the lack of renormalizability of the four-Fermion interaction
and given relatively precise predictions in the unified electroweak theory.
The discovery of these in the CERN $p \overline p$ collider
in 1983 was a dramatic confirmation of this theory.  
The gluon which mediates the color force QCD was first demonstrated
in the $e^+ e^-$ collider at DESY in Hamburg.

The minimal version of the SM, with no right-handed neutrinos and the
simplest possible ESB mechanism, has 19 arbitrary parameters: 9 fermion masses;
3 angles and one phase that specify the quark mixing matrix; 
3 gauge coupling constants; 2 parameters to specify the Higgs potential; and 
an additional phase $\theta$ that characterizes the QCD vacuum state.  
The number of parameters is larger if the ESB mechanism is more complicated or
if there are right-handed neutrinos.  Aside from
constraints imposed by renormalizability, the spectrum of elementary particles 
is also arbitrary. As discussed in Section VII, this high degree of
arbitrariness suggests that a more fundamental theory underlies the SM.

\section{Experimental establishment of the Standard Model}

The current picture of particles and interactions has been shaped and tested
by three decades of experimental studies at laboratories around the world.  We
briefly summarize here some typical and landmark results.

\subsection{Establishing QCD}

\subsubsection{Deep inelastic scattering}

Pioneering experiments at SLAC in the late 1960's directed high energy
electrons on proton and nuclear targets. The deep inelastic scattering (DIS)
process results in a deflected electron and a hadronic recoil system from the
initial baryon. The scattering occurs through the exchange of a photon coupled
to the electric charges of the participants. DIS  experiments were the
spiritual descendents of Rutherford's scattering of $\alpha$ particles by gold
atoms and, as with the earlier experiment, showed the existence of the
target's substructure. Lorentz and gauge invariance restrict the matrix
element representing the hadronic part of the interaction to two terms, each
multiplied by phenomenological form factors or structure functions.  These in
principle depend on the two independent kinematic variables; the
momentum transfer carried by the photon ($Q^{2}$) and energy loss by the
electron ($\nu$). The experiments showed that the structure functions were,
to good approximation, independent of $Q^{2}$ for fixed values of
$x=Q^{2}/2M\nu$. This `scaling' result was interpreted as evidence that the
proton contains sub-elements, originally called partons. \ The DIS scattering
occurs as the elastic scatter of the beam electron with one of the partons.
\ The original and subsequent experiments established that the struck partons
carry the fractional electric charges and half-integer spins dictated by the
quark model. \ Furthermore, the experiments demonstrated that three such
partons (valence quarks) provide the nucleon with its quantum numbers. The
variable $x$ represents the fraction of the target nucleon's momentum carried
by the struck parton, viewed in a Lorentz frame where the proton is
relativistic. The DIS experiments further showed that the charged partons
(quarks) carry only about half of the proton momentum, giving
indirect evidence for an electrically neutral partonic gluon.

\begin{figure}
\epsfxsize=8.0cm
\centerline{\epsffile{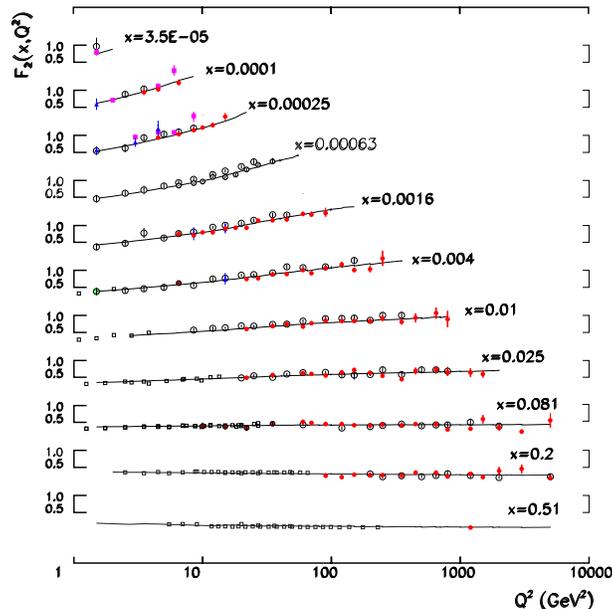}}
\caption{
The proton structure function ($F_{2}$) versus $Q^{2}$ at fixed $x$,
measured with incident electrons or muons, showing scale invariance at
larger $x$ and substantial dependence on $Q^{2}$ as $x$ becomes small. The
data are taken from the HERA $ep$ collider experiments H1 and ZEUS, as well
as the muon scattering experiments BCDMS and NMC at CERN and E665 at
FNAL.}
\label{fig:f2}
\end{figure}

\begin{figure}
\epsfxsize=8.0cm
\centerline{\epsffile{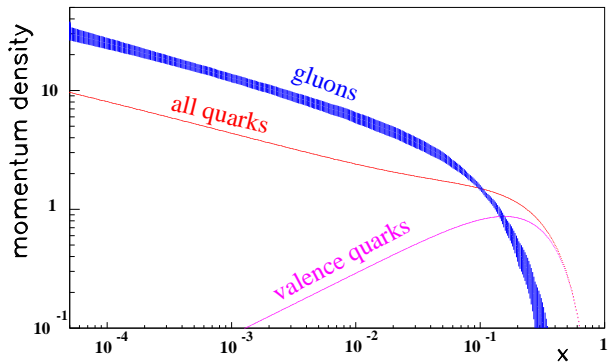}}
\vskip 3mm
\caption{
The quark and gluon momentum densities in the proton versus $x$ for 
$Q^{2}=20$ GeV$^{2}$. The integrated values of each component density gives
the
fraction of the proton momentum carried by that component. The valence $u$
and $d$ quarks carry the quantum numbers of the proton. The large number of
quarks at small $x$ arise from a `sea' of quark-antiquark pairs.  
The quark densities are from a
phenomenological fit (the CTEQ collaboration) 
to data from many sources; the gluon density
bands are the one standard deviation bounds to QCD fits to
ZEUS data (low $x$)\ and muon scattering data (higher $x$).}
\label{fig:qgdensity}
\end{figure}

Further DIS investigations using electrons, muons, and neutrinos and a variety
of targets refined this picture and demonstrated small but systematic
nonscaling behavior. The structure functions were shown to vary more rapidly
with $Q^{2}$ as $x$ decreases, in accord with the nascent QCD
prediction that the fundamental strong coupling constant $\alpha_{S}$ varies
with $Q^{2}$, and that at short distance scales (high $Q^{2}$) the number of
observable partons increases due to increasingly resolved quantum
fluctuations.  Figure~\ref{fig:f2} shows sample modern 
results for the $Q^{2}$ dependence of the dominant structure function, in
excellent accord with QCD predictions. \ The structure function {\it values}
at all $x$ depend on the quark content; the {\it increases} at larger $Q^{2}$
depend on both quark and gluon content. The data permit the
mapping of the proton's quark and gluon content exemplified in
Fig.~\ref{fig:qgdensity}.

\subsubsection{Quark and gluon jets}

\begin{figure}
\epsfxsize=8.0cm
\centerline{\epsffile{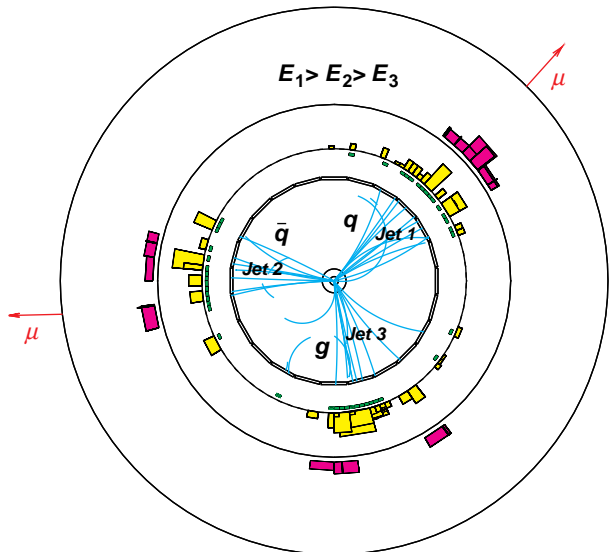}}
\vskip 1mm
\caption{A three jet event from the OPAL experiment at LEP.  The curving
tracks from the three jets may be associated with the energy deposits
in the surrounding calorimeter, shown here as histograms on the middle two
circles, 
whose bin heights are proportional to energy.   Jets 1 and 2
contain muons as indicated, suggesting that these are both quark jets
(likely from $b$ quarks).   The lowest energy jet 3 is attributed to
a radiated gluon.}
\label{fig:3jetLEP}
\end{figure}

The gluon was firmly predicted as the carrier of the color force. \ Though
its presence had been inferred because only about half the proton momentum
was found in charged constituents, direct observation of the gluon was
essential. This came from experiments at the DESY $e^{+}e^{-}$ collider
(PETRA) in 1979. The collision forms an intermediate virtual photon state,
which may subsequently decay into a pair of leptons or pair of quarks. The
colored quarks cannot emerge  intact from the collision region; instead they
create many quark-antiquark pairs from the vacuum that arrange themselves into
a set of colorless hadrons moving approximately in the directions of the
original quarks. These sprays of roughly collinear particles, called jets,
reflect the directions of the progenitor quarks. However, the quarks may
radiate quanta of QCD  (a gluon) prior to formation of the jets, just as
electrons radiate photons. \ If at sufficiently large angle to be
distinguished,  the gluon radiation evolves into a separate jet. Evidence was
found in the event energy-flow patterns for the  `three-pronged' jet
topologies expected for events containing a gluon.  Experiments at higher energy
$e^{+}e^{-}$ colliders illustrate this gluon radiation even better, as shown
in Fig.~\ref{fig:3jetLEP}.   
Studies in $e^+e^-$ and hadron collisions have verified the
expected QCD structure of the quark-gluon couplings, and their 
interference patterns.

\subsubsection{Strong coupling constant}

\begin{figure}
\epsfxsize=8.0cm
\centerline{\epsffile{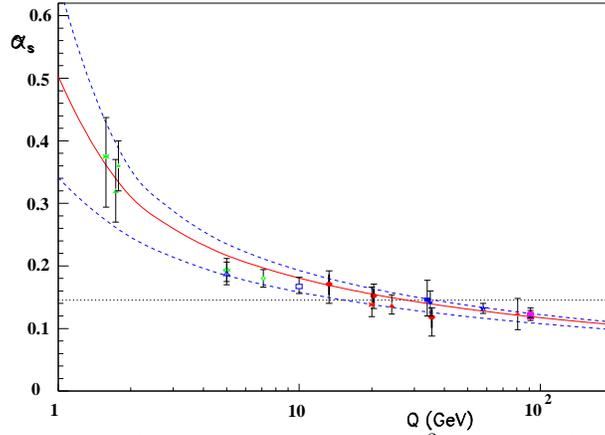}}
\caption{
The dependence of the strong coupling constant, $\alpha _{S}$, versus $Q^{2}$
using data from DIS structure
functions from $e$, $\mu$, and $\nu$ beam experiments as well as
$ep$ collider experiments, production rates of jets, heavy quark
flavors,  photons, and weak vector bosons in $e p$,
$e^+e^-$, and $p \overline p$ experiments. The data are in clear
disagreement with a strong coupling independent of
$Q^{2}$ (horizontal line).  All data agree with the dependence expected in
QCD. 
The curves correspond
to next-to-leading order calculations of the strong
coupling constanta evaluated using values
of $\alpha_{S}(M_{Z})$ of 0.1048, 0.1175 and 0.1240.}
\label{fig:alphas}
\end{figure}

The fundamental characteristic of QCD is asymptotic freedom, dictating that
the coupling constant for color interactions decreases logarithmically as
$Q^{2}$ increases. The coupling $\alpha_{S}$ can be measured in a variety of
strong interaction reactions at different $Q^{2}$ scales. At low $Q^{2}$,
processes like DIS, tau decays to hadrons, and the annihilation rate for
$e^{+}e^{-}$ into multi-hadron final states give accurate determinations of
$\alpha_{S}$. The decays of the $\Upsilon$ into three jets primarily involve
gluons, and the rate for this decay gives $\alpha_{S}$(M$_{\Upsilon}^{2}$). At
higher $Q^{2}$, studies of the $W$ and $Z$ bosons (for example, the decay
width of the $Z$, or the fraction of $W$ bosons associated with jets) measure
$\alpha_{S}$ at the 100 GeV scale. These and many other determinations have
now solidified the experimental evidence that $\alpha_{S}$ does indeed `run'
with $Q^{2}$ as expected in QCD.  
Predictions for $\alpha_S(Q^2)$, relative to its value at some
reference scale, can be made within perturbative QCD.  The current
information from many sources are compared with calculated values in 
Fig.~\ref{fig:alphas}.

\subsubsection{Strong interaction scattering of partons}

\begin{figure}
\epsfxsize=8.0cm
\centerline{\epsffile{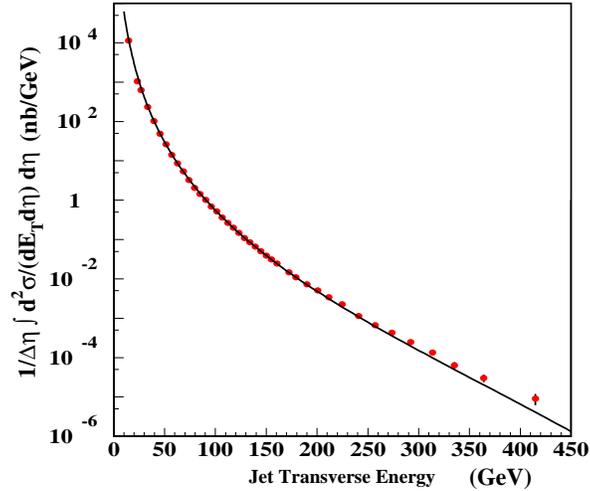}}
\caption{Inclusive jet cross section {\sl vs.} jet transverse
momentum.   The data points are from the CDF experiment.  The
curve gives the prediction of NLO QCD. }
\label{fig:highETjets}
\end{figure}

At sufficiently large $Q^{2}$ where $\alpha_{S}$ is small, the QCD
perturbation series converges sufficiently rapidly to permit  accurate
predictions.  An important process probing the highest accessible $Q^{2}$
scales is the scattering of two constituent partons (quarks or gluons) within
colliding protons and antiprotons. Figure~\ref{fig:highETjets} shows the
impressive data for the inclusive production of jets due to scattered partons
in $p\overline{p}$ collisions at 1800 GeV.  The QCD NLO predictions give
agreement with the data over nine orders of magnitude in
the cross-section.

\begin{figure}
\epsfxsize=8.0cm
\centerline{\epsffile{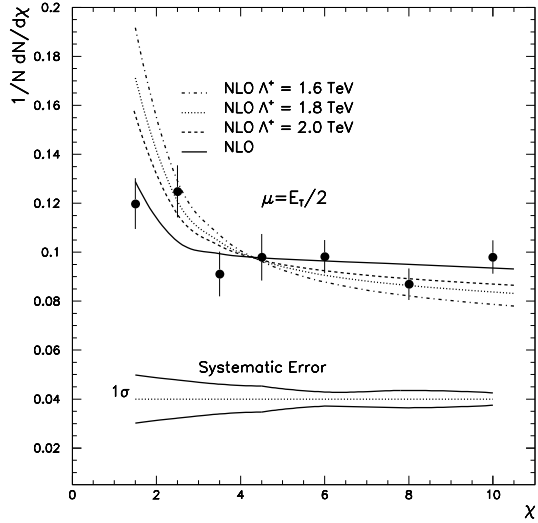}}
\caption{The dijet angular distribution from the D\O\ experiment 
plotted as a function of
$\chi$ (see text)
for which Rutherford scattering would give $d\sigma /d\chi =$ constant. 
The predictions of NLO QCD (at scale $\mu=E_T/2$) are shown by the curves.
$\Lambda$ is the compositeness scale for quark/gluon substructure, with
$\Lambda=\infty$ for no compositness (solid curve);
the data rule out values of $\Lambda < 2$ TeV.}
\label{fig:dsigdchi}
\end{figure}

The angular distribution of the two highest transverse momentum jets from
$p\overline{p}$ collisions reveals the structure of the scattering
matrix element. These amplitudes are dominated by the exchange of the spin 1
gluon. \ If this scattering were identical to Rutherford scattering, the
angular variable
$\chi=(1+|\rm{cos}\theta_{\rm{cm}}|)/(1-|\rm{cos}%
\theta_{\rm{cm}}|)$  would provide $d\sigma/d\chi$ = constant. The data
shown in Fig.~\ref{fig:dsigdchi} for dijet production show that the 
spin-1 exchange process is dominant, with clearly visible differences required
by QCD, including the varying $\alpha_{S}$. This data also demonstrates the
{\it absence} of further substructure (of the partons) to distance scales
approaching $10^{-19}$ m.

Many other measurements test the correctness of QCD in the perturbative
regime. Production of photons and $W$ and $Z$ bosons occurring in hadron
collisions are well described by QCD. Production of heavy quark pairs, such as
$t\overline{t}$, is sensitive not only to to perturbative processes, but
reflects additional effects due to multiple gluon radiation from the
scattering quarks. Within the limited statistics of current data samples,
the top quark production cross section is also in good agreement with QCD.

\subsubsection{Nonperturbative QCD}

Many physicists believe that QCD is a theory `solved in principle'. The basic
validity of QCD at large $Q^{2}$ where the coupling is small has been verified
in many experimental studies, but the large coupling at low $Q^{2}$ makes
calculation  exceedingly difficult. This low $Q^{2}$ region of QCD is relevant
to the wealth of experimental data on the static properties of nucleons, most
hadronic interactions, hadronic weak decays, nucleon and nucleus structure,
proton and neutron spin structure, and systems of hadronic matter with very
high temperature and energy densities. The ability of theory to predict such
phenomena has yet to match the experimental progress.

Several techniques for dealing with nonperturbative QCD have been developed.
The most successful address processes in which some energy or mass in the
problem is large. An example is the confrontation of data on the rates of
mesons containing heavy quarks ($c$ or $b$) decaying into lighter
hadrons, where the heavy quark can be treated nonrelativistically and its
contribution to the matrix element is taken from experiment. With this
phenomenological input, the ratios of calculated partial decay rates agree
well with experiment. Calculations based on evaluation at discrete space-time
points on a lattice and extrapolated to zero spacing have also had some
success. With computing advances and new calculational algorithms, the lattice
calculations are now advanced to the stage of calculating hadronic masses, the
strong coupling constant, and decay widths to within roughly 10 -- 20\% of the
experimental values.

The quark and gluon content of protons are consequences of QCD, much as the
wave functions of electrons in atoms are consequences of electromagnetism.
Such calculations require nonperturbative techniques. Measurements of the
small-$x$ proton structure functions at the HERA\ $ep$ collider show a much
larger increase of parton density with decreasing $x$ than were extrapolated
from larger $x$ measurements. It was also found that a large fraction
($\sim10\%$) of such events contained a final state proton essentially intact
after collision. These were called `rapidity gap' events because they were
characterized by a large interval of polar angle (or rapidity) in which no
hadrons were created between the emerging nucleon and the jet. More typical
events contain hadrons in this gap due to the exchange of the color charge
between the struck quark and the remnant quarks of the proton. Similar
phenomena have also been seen in hadron-hadron and photon-hadron scattering
processes. Calculations which analytically resum whole categories of higher
order subprocesses have been performed. In such schemes, the agent for the
elastic or quasi-elastic scattering processes is termed the `Pomeron', a
concept from the Regge theory of a previous era, now viewed as a colorless
conglomerate of colored gluons. These ideas have provided semi-quantitative
agreement with data coming from the $ep$ collider at DESY and the Tevatron.

\subsection{Establishing the Electroweak interaction}
\subsubsection{Neutral currents in neutrino scattering}

\begin{figure}
\epsfxsize=8.0cm
\centerline{\epsffile{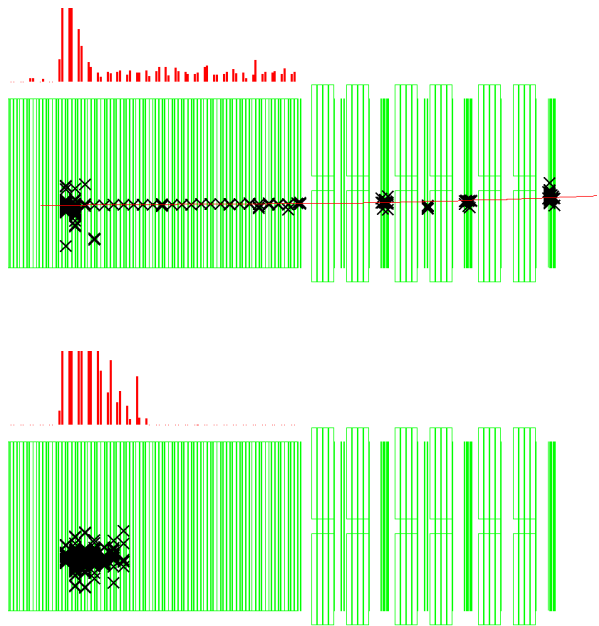}}
\caption{
Displays of events created by $\nu_\mu$'s in the CCFR experiment at
Fermilab. The upper (lower) picture is a CC (NC) interaction.  
In each case, the $\nu$ enters
from the left and interacts after traversing about 1 m of steel. The
CC event contains a visible energetic $\mu$ which penetrates
more than 10 m of steel; the NC event contains
an energetic final state $\nu$ which passes through the remainder of the
apparatus without trace. Each ($\times $) records a hit in the 
sampling planes, and the histogram above the display shows the energy
deposition in the scintillator planes interspersed in the
steel. The energy near the interaction vertex results from produced hadrons.}
\label{fig:ncccevnts}
\end{figure}

Though the electroweak theory had been proposed by 1968, it received
little experimental attention until early the next decade, when it was shown
that all such gauge theories are renormalizable. The
electroweak theory specifically proposed a new NC weak
interaction.

For virtually any scattering or decay process in which a photon might be
exchanged, the NC interaction required added Feynman diagrams
with $Z$ exchange. This predicted modifications to
known processes at very small levels. However, $Z$-exchange is the only 
mechanism by which an electrically neutral neutrino can scatter elastically 
from a quark or from an electron, leaving a neutrino in the final state.  The
theory predicted a substantial rate for this previously unanticipated
$\nu$-induced NC process. The only competitive interactions were the
well-known charged-current (CC) processes with exchange of a $W$ and a
charged final state lepton.

The NC interactions were first seen at CERN in 1973 with scattering from
nuclei at rates about 30\% of the CC scattering (as well as hints
of a purely leptonic neutrino interaction with electrons). The results were
initially treated with skepticism, since similar experiments had determined
limits close to and even below the observed signal, and other
contemporary experiments at higher energy obtained results which were
initially ambiguous. By 1974, positive and unambiguous results at FNAL
had corroborated the existence of the NC reaction using high
energy $\nu$'s. In subsequent FNAL and
CERN measurements using $\bar\nu$'s as well as $\nu$'s, the value of $\rho $ was
determined to be near unity, and the value of the weak angle, 
$\sin ^{2}\theta _{w}$,
was established. With time, the values of these parameters have been
measured more and more accurately, at low and high energies,
in $\nu$ reactions with electrons as
well as with quarks.
All are consistent with the electroweak theory and with a
single value of $\sin ^{2}\theta _{w}$.  Figure~\ref{fig:ncccevnts}
shows the characteristics of these CC and NC events.

\subsubsection{Photon and Z Interference}

\begin{figure}
\epsfxsize=8.0cm
\centerline{\epsffile{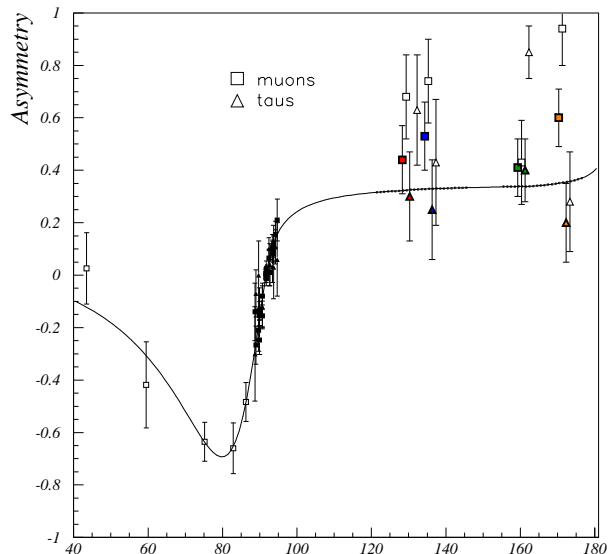}}
\caption{Forward-backward asymmetry in $e^+e^- \rightarrow \mu^+\mu^-$
and $e^+e^- \rightarrow \tau^+\tau^-$ 
as a function of energy from the DELPHI experiment at LEP.  The interference
of $\gamma$ and $Z$ contributions gives the asymmetry variation with energy,
as indicated by the SM curve.}
\label{fig:afb}
\end{figure}

The NC was found at about the anticipated level in
several different neutrino reactions, but further verification 
of the NC properties were sought.  
Though reactions of charged leptons are dominated by
photon exchange at accessible fixed target energies, the parity-violating
nature of the small $Z$-exchange contribution permits very sensitive
experimental tests. The vector part of the NC amplitude interferes
constructively or destructively with the dominant electromagnetic amplitude. 
In 1978, the first successful such effort was reported, using the
polarized electron beam at SLAC to measure the scattering asymmetry between
right-handed and left-handed beam electrons. Asymmetries of about $10^{-4}$
were observed, using several different energies, implying a single value of 
$\sin ^{2}\theta _{w}$, in agreement with neutrino measurements.

High energy $e^+ e^-$ collisions provided another
important opportunity to observe $\gamma - Z$ interference.  
By 1983 several experiments at DESY had
observed the electromagnetic-weak interference in processes where the
$e^-$ and $e^+$ annihilate to produce a final state $\mu$ pair or $\tau$
pair. 
The asymmetry grows rapidly above cm energy of 30 GeV, then changes sign
as the energy crosses the $Z$ resonance.
The weak-electromagnetic interference is beautifully confirmed in the 
LEP data as shown in  Fig.~\ref{fig:afb}.

\subsubsection{W and Z Discovery}

\begin{figure}
\epsfxsize=8.0cm
\centerline{\epsffile{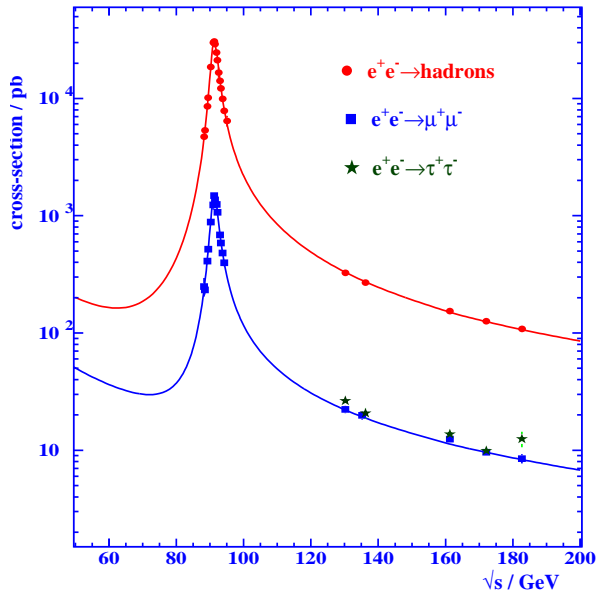}}
\caption{Dielectron invariant mass distribution for
$ee \rightarrow$ hadrons and $ee \rightarrow \mu\mu$  
from the LEP collider experiments.  The prominent
$Z$ resonance is clearly apparent.}
\label{fig:zmass}
\end{figure}

With the corroborations of the electroweak theory with $\rho \sim 1$, and
several consistent measurements of the one undetermined parameter, $\sin
^{2}\theta _{w}$, reliable predictions existed by 1980 for the masses of the
vector bosons, $W$ and $Z$.  The predicted masses, 
about 80 and 90 GeV respectively, 
were not accessible to $e^+e^-$ colliders or
fixed target experiments, but adequate cm energy was possible
with existing proton accelerators, so long as the collisions were between
two such beams. Unfortunately, none had the two rings required to collide
protons with protons.

\begin{figure}
\epsfxsize=8.0cm
\centerline{\epsffile{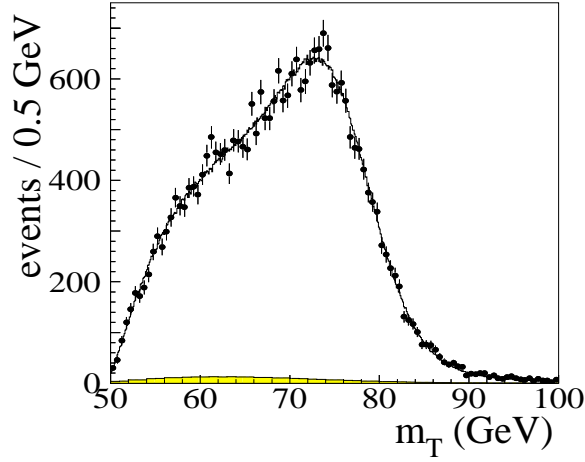}}
\caption{Transverse mass distribution for $W \rightarrow e \nu$ from the
D\O\ experiment.   
The transverse mass is defined as $M_T =$$(2E_T^e E_T^\nu$
$(1-\cos\phi^{e\nu}))^{1/2}$ with $E_T^e$ and $E_T^\nu$ the transverse
energies of electron and neutrino and $\phi^{e\nu}$ the azimuthal angle between
them.   $M_T$ has its Jacobian edge at the mass of the $W$ boson.}
\label{fig:wmass}
\end{figure}

A concerted effort was mounted at CERN to find the predicted bosons. To save
the cost and time of building a second accelerating ring, systems were
constructed to produce and accumulate large numbers of antiprotons, gather
these and `cool' them into a beam, and then accelerate them in the
existing accelerator to collide with a similar beam of protons. In 1983, the
$W$ and $Z$ decays were observed with the anticipated
masses.    Present-day measurements from LEP
(Fig.~\ref{fig:zmass})
give a fractional $Z$ mass 
precision of about $10^{-5}$ and studies at the FNAL $p \overline p$
collider give a fractional $W$ mass precision of about $10^{-3}$
(Fig.~\ref{fig:wmass}).

\begin{figure}
\epsfxsize=8.0cm
\centerline{\epsffile{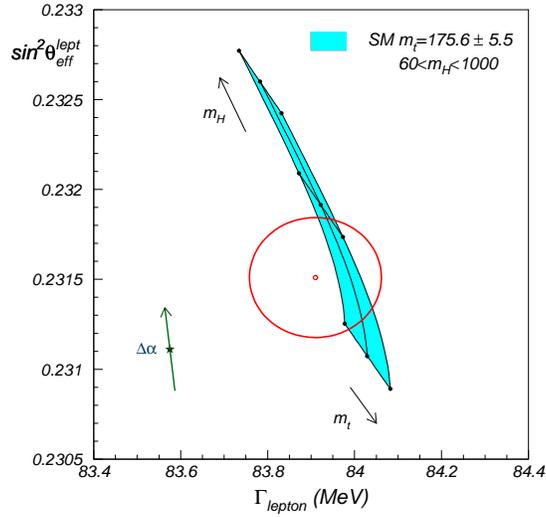}}
\caption{The allowed region for $\sin^2\theta_w$ {\sl vs.} 
$\Gamma_{\rm lepton}$ in the context of the SM, showing the need for the higher
order EW corrections.   The region within the ellipse is allowed (at 1 standard
deviation)
by the many precision measurements at the LEP and SLC $ee$ colliders and the
FNAL $p \overline p$ collider;
the shaded region comes from the measurements of the top mass at FNAL, for
a range of possible Higgs masses.
The star, well outside the allowed region,
gives the expected value in the SM without
the higher order EW corrections.}
\label{fig:sthwgamma}
\end{figure}

\subsubsection{Z Properties and precision tests of the electroweak SM}

The LEP and SLAC Linear Collider experiments have made 
many precise measurements of
the properties of the $Z$, refining and testing the electroweak model.
The asymmetries due to weak-electromagnetic interference discussed above
were extended to include all lepton species, $c$- and $b$-quark
pairs, and light-quark pairs, as well as polarization 
asymmetries involving $\tau$ pairs, and initial state 
left- or right-handed electrons.   From these data, the underlying vector and
axial couplings to fermions have been extracted and found to be in 
excellent agreement with the SM, and with lepton universality.
The fundamental weak mixing parameter, sin$^2\theta_w$, has been determined
from these and other inputs to be 0.23152 $\pm$ 0.00023.

The total width of the $Z$ is determined to be 
2.4948 $\pm$ 0.0025 GeV; the invisible decay contributions to this
total width allow the number of light ($m_\nu < m_Z/2$) neutrino generations
to be measured: $N_\nu = 2.993 \pm 0.011$, confirming another aspect of the
SM.  The partial widths for the $Z$ were measured, again testing
the SM to the few percent level, and restricting possible additional
non-SM particle contributions to the quantum loop 
corrections.  The electroweak and
QCD higher order corrections modify the expectations for all observables.
Figure ~\ref{fig:sthwgamma} 
shows the allowed values
in the sin$^2\theta_w$ {\it vs.} $\Gamma_{\rm lepton}$ plane
under the assumption that the SM is valid.
Even accounting for uncertainties in the Higgs boson mass, it is
clear that the higher order electroweak corrections are required.

\begin{figure}
\epsfxsize=9.0cm
\centerline{\epsffile{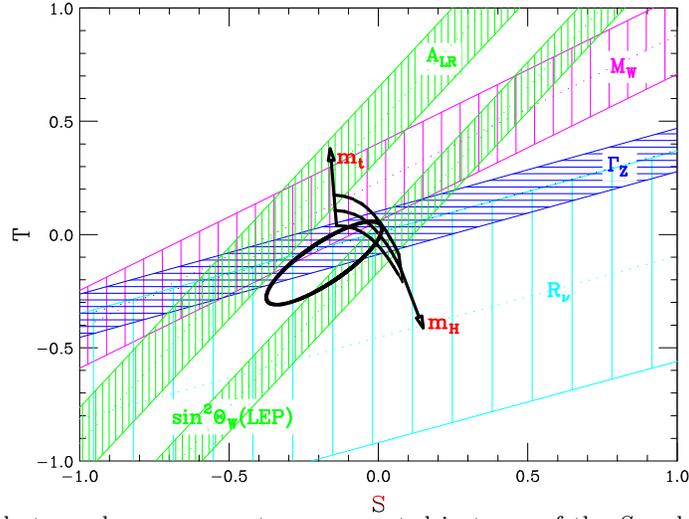}}
\caption{
Several precise electroweak 
measurements are presented
in terms of the $S$ and $T$ variables which characterize 
the consistency of observables with the SM.
 The bands shown from the experimental measurements of $A_{\rm LR}$ (SLC),
$\Gamma_Z$ (LEP), sin$^2\theta_w$ (LEP), $M_W$ (FNAL and CERN) 
and $R_\nu$ ($\nu$ deep inelastic scattering
experiments at CERN and FNAL)
indicate the allowed regions in $S$, $T$ space.
The half-chevron region centered on $S=T=0$ gives the 
prediction for top mass = $175.5 \pm 5.5$ GeV
and Higgs mass between 70 and 1000 GeV, providing the SM is correct. 
A fit to all electroweak data
yields the 68\% confidence region bounded by the ellipse and shows the
consistency of the data and the agreement with the minimal SM theory. }
\label{fig:ewst}
\end{figure}

Taken together, the body of electroweak observables 
tests the overall consistency of the SM.   
Extensions of the SM would result in modification of observables
at quantum loop level; dominant non-SM effects should modify the 
vacuum polarization terms, and may be parametrized in terms of 
weak-isospin conserving ($S$) and weak-isospin breaking ($T$) couplings.
$S$ and $T$ may be chosen to be zero for specific top quark and Higgs mass
values in the minimal SM;  
Fig.~\ref{fig:ewst} 
shows the constraints afforded by several
precision measurements, and indicates the level to which extensions
to the SM are ruled out.

\subsubsection{The top quark}

\begin{figure}
\epsfxsize=8.0cm
\centerline{\epsffile{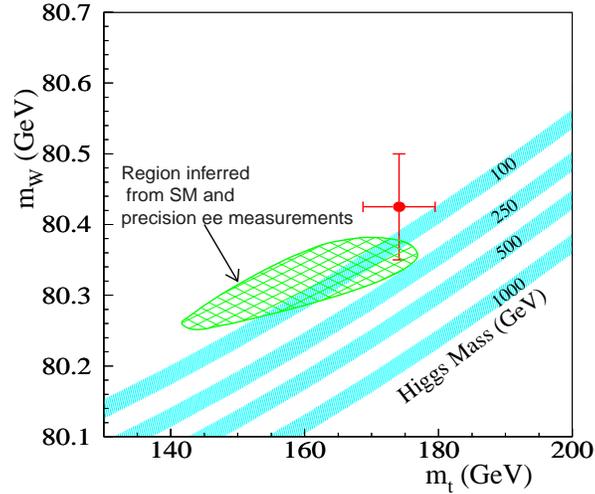}}
\caption{
$W$ boson mass {\sl vs.} top quark mass.
The data point is the average of FNAL data for the top quark mass 
and FNAL and CERN data for the $W$ boson mass.
The shaded bands
give the expected values for specific conventional
Higgs boson mass values in the context of the minimal SM.
The cross-hatched region shows the predictions for $m_W$ and $m_{\rm top}$,
at 68\% confidence level, from precision electroweak measurements of $Z$ boson
properties.}
\label{fig:mtmw}
\end{figure}

The top quark was expected even before measurements in $e^+ e^-$ 
scattering unambiguously determined the $b$ quark to be the $I_3=-{1\over2}$
member of an isospin doublet.   In 1995, the two FNAL $p \overline p$
collider experiments reported the first observations of the top.
Though expected as the last fermion in the SM, its mass of about
175 GeV is startlingly large compared to its companion $b$, at about 4.5 GeV,
and to all other fermion masses. 
The $t$ decays nearly always into
a $W$ and a $b$, with final states governed by the subsequent decay of
the $W$.   The large top quark mass gives it the largest 
fermionic coupling to the Higgs sector.  Since its mass is of order the Higgs
vacuum expectation value $<$$|\phi|$$>$, it is possible that
the top plays a unique role in ESB.  The top quark mass is
now measured with precision of about 3\%.  Together with
other precision electroweak determinations, the mass gives useful SM 
contraints on the unknown Higgs boson mass,
as shown in Fig.~\ref{fig:mtmw}.
At present, measurements require a SM Higgs boson mass 
less than 420 GeV at 95\% confidence level.  Such constraints place
the Higgs boson, 
if it exists, within the range of anticipated experiments. 

\subsubsection{Trilinear Gauge Couplings}

\begin{figure}
\epsfxsize=8.0cm
\centerline{\epsffile{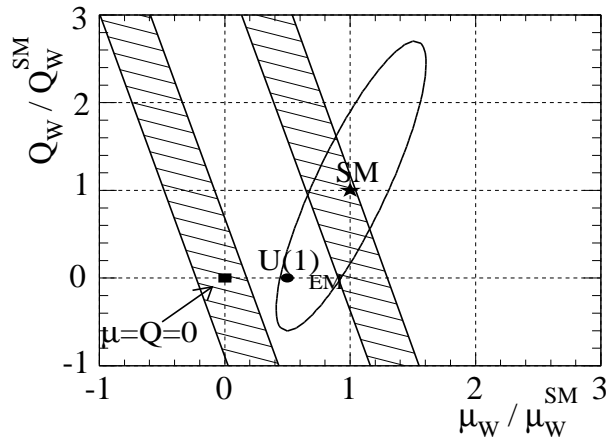}}
\caption{The $W$ boson electric quadrupole moment
{\sl vs} magnetic dipole moment from $W\gamma$ production
relative to their SM values.  The ellipse shows the 95\% confidence level
limit from the D\O\ experiment with both $Q$ and $\mu$ allowed to vary.
Limits from $b \rightarrow s \gamma$ from CLEO at Cornell and ALEPH at LEP
are shown as the hatched bands.  
The star shows the
moments if the SM couplings are correct; the filled circle
labelled U(1)$_{\rm EM}$ corresponds to a SM SU(2) coupling of zero. }
\label{fig:wwgamma}
\end{figure}

The gauge symmetry of the electroweak SM exactly specifies the couplings of
the $W$, $Z$
and $\gamma$ bosons to each other.  These gauge couplings may be 
probed through the production of  boson pairs:  $WW$, $W\gamma$,
$WZ$, $Z\gamma$ and $ZZ$.   The SM specifies  precisely the 
interference terms for all these processes.  The diboson
production reactions have been observed in FNAL collider experiments
and the $WW$ production has been seen at LEP.  Limits have been placed on 
possible anomalous couplings beyond the SM.   For 
$WW\gamma$, the experiments have shown that the full
electroweak gauge structure of the SM 
is necessary, as shown in
Fig~\ref{fig:wwgamma}, and constrain the
anomalous magnetic dipole
and electric quadrupole moments of the $W$.

\subsubsection{Quark mixing matrix}

The generalization
of the rotation of the down-strange weak interaction eigenstates
from the strong interaction basis indicated in (2) to the case
of three generations
gives a $3\times 3$ unitary transformation matrix, {\bf V}, whose elements
are the mixing amplitudes among the $d$, $s$ and $b$ 
quarks.   Four parameters -- three real numbers ({\it e.g.} 
Euler angles) and one phase -- are
needed to specify this matrix.   The real elements of this
`Cabibbo-Kobayashi-Maskawa'  (CKM) matrix are determined from various
experimental
studies of weak flavor-changing interactions and decays. 
The decay rates of $c$ and $b$ quarks depend on the CKM 
elements connecting the second and third generation.  These 
have been extensively explored in $e^+e^-$ and hadronic collisions which
copiously
produce $B$ and charmed mesons at Cornell, DESY, and FNAL.
The pattern that emerges shows a hierarchy in which
the mixing between first and second generation is of order the
Cabibbo angle, $\lambda = \sin\theta_c$, those between the second and third
generation are of order $\lambda^2$ and, between first and third generation,
of order $\lambda^3$.

\begin{figure}
\epsfxsize=8.0cm
\centerline{\epsffile{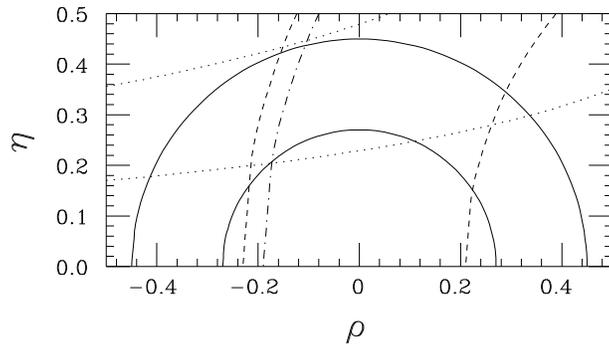}}
\caption{Experimentally allowed regions in the $\rho$ $\eta$
plane from experiments.  The region between the solid semicircles are 
from the ratio of $b$ quark decays into $u$ or $c$ quarks.
The CP violating amplitudes from $K_L^0$ decays give the band between the dotted
hyperbolae.   
The region between the dashed semicircles are allowed by
measurements of $B^0 - \overline {B}^0$ mixing.
The constraint imposed from current limits on $B_s^0 - \overline {B}_s^0$
mixing is to the right of the dot-dashed semicircle.  Current experiments thus
are consistent, and favor non-zero values of the CP-violating parameter $\eta$.
}
\label{fig:rhoeta}
\end{figure}

A non-zero CKM phase would provide CP violating effects such as the
decay $K_L^0 \rightarrow \pi \pi$, as well as different decay rates 
for $B^0$ and $\overline {B}^0$ into CP-eigenstate final states.
CP violation has only been observed to date in the neutral $K$ decays,
and is consistent with (though not requiring) the description embodied in
the CKM matrix.
Well-defined predictions of the CKM phase for a variety of $B$ decay
asymmetries will be tested in experiments at SLAC, KEK in Japan, 
Cornell, DESY and FNAL in the coming few years.   The unitarity relations
${{\bf V}^{\dag}}_{ij} {\bf V}_{jk} = \delta_{ik}$ impose constraints
on the observables that must be satisfied if CP violation is indeed
embedded in the CKM matrix and if there are but three quark generations.
Figure~\ref{fig:rhoeta} 
shows the current status of the  
constraints on the 
real and
imaginary parts ($\rho, \eta$) of the complex factor necessary
if the origins of CP violation are inherent to the CKM matrix.

\section{Unresolved issues: beyond the Standard Model}

While the SM has proven highly successful in correlating vast
amounts of data, a major aspect of it is as yet untested, namely the origin of
ESB.  The Higgs mechanism described in Section IV is
just the simplest ansatz that is compatible with observation.  It predicts the
existence of a scalar particle, but not its mass; current LEP data provide a 
lower limit: $m_H > 80$ GeV.   The Higgs mass is determined by its coupling 
constant $\lambda$ [{\it c.f.} Eq.(3)] and its vacuum value $v$: $m_H \approx
\lambda\times 348$GeV.  A Higgs mass of a TeV or more would imply strong
coupling of longitudinally polarized $W$ and $Z$ bosons that are the remnants of
the `eaten' Goldstone boson partners of the physical Higgs particle.  It can
be shown quite generally that if there is no Higgs particle with a mass less
than about a TeV, strong $W,Z$ scattering will occur at TeV cm
energies; the observation of this scattering requires multi-TeV proton-proton 
cm energies, as will be achieved at the LHC. 

However, the introduction of an elementary scalar field in QFT
is highly problematic. Its mass is subject to large quantum corrections 
that make it difficult to understand how it can be as small as a TeV or 
less in the presence of large scales in nature like the Planck scale of 
$10^{19}$ GeV or possibly a scale of coupling constant unification at 
$10^{16}$ GeV.  Moreover, a strongly interacting scalar field theory is not 
self-consistent as a fundamental theory: the coupling constant grows with 
energy and therefore any finite coupling at high energy implies a weakly 
coupled theory at low energy.  There is therefore strong reason to believe 
that the simple Higgs mechanism described in Section IV is incorrect or 
incomplete, and that ESB must be associated with
fundamentally new physics. Several possibilities for addressing these problems 
have been suggested; their common thread is the implication that the Standard
Model is an excellent low energy approximation to a more fundamental theory,
and that clues to this theory should appear at LHC energies or below.

For example, if quarks and leptons are 
composites of yet more fundamental entities, the SM is
a good approximation to nature only at energies small compared with the 
inverse radius of compositeness $\Lambda$. The observed scale of 
ESB, $v\sim {1\over4}$ TeV, might emerge 
naturally in connection with the compositeness scale.  A signature of
compositeness would be deviations from SM predictions for high 
energy scattering of quarks and leptons. Observed consistency ({\it e.g.}, 
Fig.~\ref{fig:dsigdchi}) 
with the SM provides limits on $\Lambda$ that 
are considerably higher than the scale $v$ of ESB.

Another approach seeks only to eliminate the troublesome scalars as fundamental
fields.  Indeed, the spontaneous breaking of chiral symmetry by a 
quark-antiquark condensate in QCD also contributes to ESB.  
If this were its only source, the $W,Z$ masses would be determined by
the 100 MeV scale at which QCD is strongly coupled:
$m_W = \cos\theta_w m_Z \approx$ 30 MeV.  To explain the much larger observed
masses, one postulates a new gauge interaction called technicolor that
is strongly coupled at the scale $v\sim {1\over4}$ TeV.  
At this scale fermions with 
technicolor charge condense, spontaneously breaking both a chiral symmetry 
and the electroweak gauge symmetry.  The longitudinally polarized components of
the massive $W$ and $Z$ are composite pseudoscalars that are Goldstone bosons of
the broken chiral symmetry, analogous to the pions of QCD.  This is a concrete
realization of a scenario with no light scalar particle, but with strong
$W,Z$ couplings in the TeV regime, predicting a wealth of new composite
particles with TeV masses.  However, it has proven difficult to construct
explicit models that are consistent with all data, especially
the increasingly precise measurements that probe electroweak quantum 
corrections to $W$ and $Z$ self-energies; these data 
(Figs.~\ref{fig:ewst},\ref{fig:mtmw}) 
appear to favor an elementary scalar less massive than a few hundred GeV. 

The quantum instability of elementary scalar masses can be overcome by extending
the symmetry of the theory to one that relates bosons to fermions, known as
supersymmetry.  Since quantum corrections from fermions and bosons have
opposite signs, many of them cancel in a supersymmetric
theory, and scalar masses are no more unstable than fermion
masses, whose smallness can be understood in terms of approximate chiral
symmetries. This requires doubling the number of spin degrees of freedom
for matter and gauge particles: for every fermion $f$ there is a complex scalar 
partner ${\tilde f}$ with the same internal quantum numbers, and for every 
gauge boson $v$ there is a spin-${1\over2}$ partner ${\tilde v}$.  
In addition, the cancellation of quantum
gauge anomalies and the generation of masses for all charged fermions requires
at least two distinct Higgs doublets with their fermion superpartners.
Mass limits on matter and gauge superpartners 
($m_{{\tilde\ell},{\tilde W}} >50$ GeV,$\;m_{{\tilde q},{\tilde g}} >200$ GeV) 
imply that supersymmetry is broken in nature.  However, if 
fermion-boson superpartner mass splittings are less than about a TeV, 
quantum corrections to the Higgs mass will be suppressed to the same level.
For this scenario to provide a viable explanation of the ESB scale, at least 
some superpartners must be light enough to be observed at the LHC.
 
Another untested aspect of the SM is the origin of CP violation,
conventionally introduced through complex Yukawa couplings of fermions 
to Higgs particles, resulting in complex parameters in the CKM matrix.  This 
ansatz is sufficient to explain the observed CP violation in $K$-decay, is 
consistent with limits on CP violation in other processes, and predicts
observable CP violating effects in $B$-decay.  Planned experiments at new and 
upgraded facilities capable of producing tens of millions of $B$-mesons will 
determine if this model correctly describes CP violation, at least at 
relatively low energy. A hint that some other source of CP violation may be 
needed, perhaps manifest only at higher energies, comes from 
the observed predominance of matter over anti-matter in the universe.

\begin{figure}
\epsfxsize=8.0cm
\centerline{\epsffile{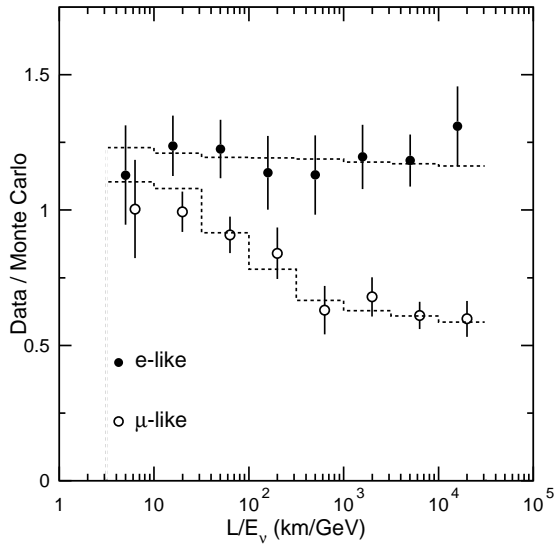}}
\vskip 6mm
\caption{
The ratio of the number of $\nu_e$ and $\nu_\mu$ interactions in the 
SuperKamiokande detector to the Monte Carlo expectations for each,
as a function of $L/E_\nu$.
$L$ is the distance of travel from neutrino production in
the earth's atmosphere and $E_\nu$ is the neutrino energy.
Neutrinos produced on the far side of the earth and going upwards in
the detector contribute at the largest $L/E_\nu$.   The Monte Carlo
curves are computed for the best fit
difference in mass squared between oscillating
neutrinos of $2.2 \times 10^{-3}$ eV$^2$ and maximal mixing.
}
\label{fig:nuosc}
\end{figure}

While in the minimal formulation of the SM, neutrinos are massless and
exist only in left-handed states, there have been persistent indirect
indications for
both neutrino masses and mixing of neutrino flavors.  Nonzero neutrino mass
and lepton flavor violation would produce spontaneous oscillation of
neutrinos from one flavor to another in a manner similar to the strangeness
oscillations of neutral K-mesons.  Solar neutrinos of energies between 0.1
to 10 MeV have been observed to arrive at the earth at a rate significantly
below predictions from solar models.  A possible interpretation is the
oscillation of $\nu_e$'s from the solar nuclear reactions to some other
species, not observable as CC interactions
in detectors due to energy conservation.  Model calculations indicate that
both solar-matter-enhanced neutrino mixing and vacuum oscillations over the
sun-earth transit distance are viable solutions.  A deficit of $\nu_\mu$
relative to $\nu_e$ from the decay products of mesons produced by cosmic
ray interactions in the atmosphere has been seen in several experiments.
Recent data from the Japan-U.S. SuperKamiokande experiment, a large water
Cerenkov detector located in Japan, corroborate this anomaly.
Furthermore, their 
observed $\nu_\mu$ and $\nu_e$ neutrino interaction rates
plotted against the relativistic distance of neutrino transit
(Fig. \ref{fig:nuosc})
provides strong
evidence for oscillation of $\nu_\mu$ into $\nu_\tau$ -- or into an unseen
``sterile'' neutrino.  An experimental anomaly observed at Los Alamos
involves an observation of $\nu_e$ interactions from a beam of $\nu_\mu$.
These indications of neutrino oscillations are spurring 
efforts worldwide to resolve the patterns of flavor oscillations of massive
neutrinos. 

The origins of ESB and of CP violation, as well as the issue of the neutrino
mass, are unfinished aspects of the SM. However, the very structure of the 
SM raises many further questions, strongly indicating that this model provides 
an incomplete description of the elementary structure of nature.

\begin{figure}
\epsfxsize=8.0cm
\centerline{\epsffile{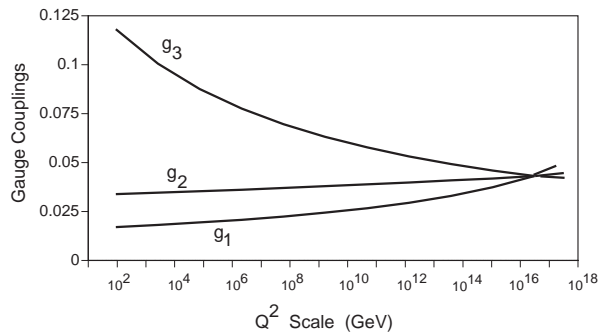}}
\caption{Gauge couplings $g_1, g_2, g_3$ as a function of 
$q^2$ in the context of the minimal supersymmetric model, showing
unification around $10^{16}$ GeV.}
\label{fig:unification}
\end{figure}

The SM is characterized by a large number of parameters. As noted 
above, three of these -- the gauge coupling constants -- approximately
unify at a scale of about 10$^{16}$ GeV.  
In fact, when the coupling evolution is
calculated using only the content of the SM, unification is not
precisely achieved at a single point: an exact realization of coupling
unification requires new particles beyond those in the SM spectrum. 
It is tantalizing that exact unification can be achieved with the
particle content of the minimal supersymmetric extension of the SM 
if superpartner masses lie in a range between 100 GeV and 10 TeV 
(Fig. \ref{fig:unification}).  

Coupling unification, if true, provides compelling evidence 
that, above the scale of unification, physics is described
by a more fundamental theory incorporating the SM interactions
in a fully unified way.  One possibility, Grand Unified Theory (GUT), invokes a 
larger gauge group, characterized by a single coupling constant, that is broken
 to the SM gauge group by a Higgs vacuum value, $v\sim 10^{16}$ GeV.
Couplings differ at low energies because some particles acquire large masses 
from this Higgs field; symmetry is restored at energy scales above $10^{16}$ GeV
where these masses are unimportant. 
Another possibility is that a completely different theory emerges above the
scale of unification, such as a superstring theory in ten dimensional 
space-time -- perhaps itself an approximation to a yet more fundamental theory
in eleven dimensions (see the following article). 
In string-derived models, 
coupling unification near the string scale is due to the fact that all gauge 
coupling constants are determined by the vacuum value of a single scalar 
field.

Most of the remaining parameters of the SM, namely the fermion masses 
and the elements of the CKM matrix (including a CP violating phase) are
governed by Yukawa couplings of fermions to the Higgs fields.  The
observed hierarchies among quark fermion masses and mixing parameters are
strongly suggestive that new physics must be at play here as well.  If there are
no right-handed neutrinos, the
SM, with its minimal Higgs content, naturally explains the absence,
or very strong suppression, of neutrino masses.  However many extensions of
the SM, including GUT and string-derived models,
require right-handed neutrinos, in which case additional new physics is 
needed to account for the extreme smallness of neutrino masses.

Many models have been proposed in attempts to understand the observed patterns 
of fermion masses and mixing.  These include extended gauge or global
symmetries, some in the context of GUT or string theory, as well 
as the possibility of quark and lepton compositeness.  Unlike the issues of 
ESB and CP violation, there is no well-defined 
energy scale or set of experiments that is certain to provide positive
clues, but these questions can be attacked on a variety of fronts.  These
include precision measurements of the CKM matrix elements, searches for 
flavor-changing transitions that are forbidden in the SM, and
high energy searches for new particles such as new gauge bosons or
excited states of quarks and leptons.  

The SM has another parameter, $\theta$, that governs the strength 
of CP violation induced by nonperturbative effects in QCD.  The experimental
limit on the neutron electric dipole moment imposes the constraint 
$\theta < 10^{-9}$, again suggestive of an additional symmetry that is not
manifest in the SM. Many other questions remain unresolved; 
some have profound implications for cosmology, discussed in Chapter 5. 
Is the left/right asymmetry of the electroweak interaction
a fundamental property of nature, or is mirror symmetry restored at high 
energy? Is the proton stable?  GUT extensions of the SM 
generally predict proton decay at some level, mediated by bosons that carry both
quark and lepton numbers.  Why are there three families of matter? Some 
suggested answers invoke extended symmetries; others conjecture fermion 
compositeness; in string theory the particle spectrum of the low energy theory 
is determined by the topology of the compact manifold of additional 
spatial dimensions.  Why is the cosmological constant so tiny, when, in the
context of QFT, one would expect its scale to be governed by other scales in the
theory, such as the ESB scale of a TeV, or the Planck scale of $10^{19}$ GeV?
The SM is incomplete in that it does not incorporate gravity.  Superstrings or
membranes, the only candidates at present for a quantum
theory of gravity, embed the SM in a larger theory whose full
content cannot be predicted at present, but which is expected to include a 
rich spectrum of new particles at higher energies.

Future experiments can severely constrain possible extensions of the Standard 
Model, and the discovery of unanticipated new phenomena may provide a 
powerful window into a more fundamental description of nature.

Thousands of original papers have contributed to the evolution of the Standard
Model.  
We apologize for omitting references to these, and for the necessarily
incomplete coverage of many incisive results.  We offer some recent 
reviews which give an entry into this illuminating and impressive literature.

\end{document}